\documentclass[twoside,twocolumn,9pt]{article}
\usepackage{extsizes}
\usepackage[super,sort&compress,comma]{natbib} 
\usepackage[version=3]{mhchem}
\usepackage[left=1.5cm, right=1.5cm, top=1.785cm, bottom=2.0cm]{geometry}
\usepackage{balance}
\usepackage{mathptmx}
\usepackage{sectsty}
\usepackage{graphicx} 
\usepackage{lastpage}
\usepackage[format=plain,justification=justified,singlelinecheck=false,font={stretch=1.125,small,sf},labelfont=bf,labelsep=space]{caption}
\usepackage{float}
\usepackage{fancyhdr}
\usepackage{fnpos}
\usepackage[english]{babel}
\usepackage{upgreek}
\usepackage{graphicx}
\usepackage{amsmath}

\addto{\captionsenglish}{%
  
}
\usepackage{array}
\usepackage{droidsans}
\usepackage{charter}
\usepackage[T1]{fontenc}
\usepackage[usenames,dvipsnames]{xcolor}
\usepackage{setspace}
\usepackage[compact]{titlesec}
\usepackage{hyperref}
\usepackage{epstopdf}
\definecolor{cream}{RGB}{222,217,201}

\begin{document}

\pagestyle{fancy}
\thispagestyle{plain}
\fancypagestyle{plain}{
\renewcommand{\headrulewidth}{0pt}
}

\makeFNbottom
\makeatletter
\renewcommand\LARGE{\@setfontsize\LARGE{15pt}{17}}
\renewcommand\Large{\@setfontsize\Large{12pt}{14}}
\renewcommand\large{\@setfontsize\large{10pt}{12}}
\renewcommand\footnotesize{\@setfontsize\footnotesize{7pt}{10}}
\renewcommand\scriptsize{\@setfontsize\scriptsize{7pt}{7}}
\makeatother

\renewcommand{\thefootnote}{\fnsymbol{footnote}}
\renewcommand\footnoterule{\vspace*{1pt}%
\color{cream}\hrule width 3.5in height 0.4pt \color{black} \vspace*{5pt}} 
\setcounter{secnumdepth}{5}

\makeatletter 
\renewcommand\@biblabel[1]{#1}            
\renewcommand\@makefntext[1]%
{\noindent\makebox[0pt][r]{\@thefnmark\,}#1}
\makeatother 
\renewcommand{\figurename}{\small{Fig.}~}
\sectionfont{\sffamily\Large}
\subsectionfont{\normalsize}
\subsubsectionfont{\bf}
\setstretch{1.125} 
\setlength{\skip\footins}{0.8cm}
\setlength{\footnotesep}{0.25cm}
\setlength{\jot}{10pt}
\titlespacing*{\section}{0pt}{4pt}{4pt}
\titlespacing*{\subsection}{0pt}{15pt}{1pt}

\fancyfoot{}
\fancyfoot[RO]{\footnotesize{\sffamily{1--\pageref{LastPage} ~\textbar  \hspace{2pt}\thepage}}}
\fancyfoot[LE]{\footnotesize{\sffamily{\thepage~\textbar\hspace{3.45cm} 1--\pageref{LastPage}}}}
\fancyhead{}
\renewcommand{\headrulewidth}{0pt} 
\renewcommand{\footrulewidth}{0pt}
\setlength{\arrayrulewidth}{1pt}
\setlength{\columnsep}{6.5mm}
\setlength\bibsep{1pt}

\makeatletter 
\newlength{\figrulesep} 
\setlength{\figrulesep}{0.5\textfloatsep} 

\newcommand{\topfigrule}{\vspace*{-1pt}%
\noindent{\color{cream}\rule[-\figrulesep]{\columnwidth}{1.5pt}} }

\newcommand{\botfigrule}{\vspace*{-2pt}%
\noindent{\color{cream}\rule[\figrulesep]{\columnwidth}{1.5pt}} }

\newcommand{\dblfigrule}{\vspace*{-1pt}%
\noindent{\color{cream}\rule[-\figrulesep]{\textwidth}{1.5pt}} }

\makeatother
\twocolumn[
  \begin{@twocolumnfalse}

\vspace{1em}
\sffamily
\LARGE{\textbf{Electrical properties of highly-doped MBE-grown gallium phosphide nanowires}} \\

\noindent\large{Vladislav Sharov,$^{\ast}$\textit{$^{a,b}$} Kristina Novikova,\textit{$^{a,c}$} Alexey Mozharov\textit{$^{a,c}$}, Vladimir Fedorov\textit{$^{a,c}$},Prokhor Alekseev\textit{$^{b}$},Ivan Mukhin\textit{$^{a,c,d}$}} \\

 \end{@twocolumnfalse} \vspace{0.6cm}

  ]
\renewcommand*\rmdefault{bch}\normalfont\upshape
\rmfamily
\section*{}
\vspace{-1cm}

\footnotetext{\textit{$^{a}$~Alferov University, Saint-Petersburg 194021, Russia}}
\footnotetext{\textit{$^{b}$~Ioffe Institute, Saint-Petersburg 194021, Russia}}
\footnotetext{\textit{$^{c}$~Peter the Great Saint-Petersburg Polytechnic University, Saint-Petersburg 195251, Russia}}
\footnotetext{\textit{$^{d}$~Saint Petersburg State University, Saint Petersburg 199034, Russia}}
\footnotetext{\textit{$^{\ast}$~E-mail: vl\_sharov@mail.ru}}

\sffamily{\textbf{Efficient doping of semiconductor nanowires remains a major challenge towards the commercialization of nanowire-based devices. In this work we investigate the growth regimes and electrical properties of MBE-grown p- and n-type gallium phosphide nanowires doped with Be and Si respectively. Electrical conductivity of individual nanowires is quantitatively studied via atomic force microscopy supported with numerical analysis. Based on conductivity measurements, we provide growth strategies for achieving the doping level up to N\textsubscript{D}=5$\cdot$10\textsuperscript{18} cm\textsuperscript{-3} and N\textsubscript{A}=2$\cdot$10\textsuperscript{19} cm\textsuperscript{-3} for GaP:Si and GaP:Be nanowires respectively, which is high enough to be demanded for technological applications.}}\\

\rmfamily 

\section{Introduction}
Semiconductor nanowires (NWs) provide new demanded technological possibilities including the direct monolithic integration of III-V materials on silicon, reduced material consumption in photovoltaic and photo-detecting applications without performance degradation, synthesis of new materials and nanostructures such as wurtzite III-V and crystal phase quantum dots as well as new platform for semiconductor straintronics due to high mechanical strength \cite{barrigon2019synthesis,mcintyre2020semiconductor,standing2015efficient,wilson2020integrated,trofimov2020perovskite,khmelevskaia2021directly,sharov2022nanoscale,kuznetsov2023elastic,kim2021doping}. In particular, gallium phosphide nanowires (GaP NWs), which possess high refractive index and broad transparency window, are promising for nanophotonic wave-guides and transparent emitters in axially-heterostructured LEDs, lasers and single photon sources a well as for photoelectrochemical cells \cite{standing2015efficient,wilson2020integrated,trofimov2020perovskite,khmelevskaia2021directly,sharov2022nanoscale,kuznetsov2023elastic}.

Despite the extensive development of epitaxial NW growth techniques for the last decades, their device implementation is still limited. To date, successful and controllable NW doping is considered to be the main unsolved issue towards their commercialization \cite{kim2021doping}. The control of electrically active dopant concentrations and concentration depth profiles is known to be the key for tuning the properties of semiconductor material and fabricating semiconductor devices. Doping control is challenging as the NW growth conditions do not always favor the desirable dopant incorporation. It was shown that dopant concentration can be restricted by the low solubility in the catalytic droplet during NW vapor-liquid-solid (VLS) growth \cite{dubrovskii2020te}. Vice versa, the excessive dopant flux can affect the growth mechanism of NWs, their structure and morphology \cite{diak2023ultrathin, Goktas_2020_twin}. In the case of amphoteric impurities, such as Si, self-compensation phenomena are often observed \cite{dubrovskii2020_GaAs_Si}. Thus, the development of the effective approaches for both n- and p-type doping of epitaxial GaP NWs is a crucial step for their device implementation.\\
To date, there are few research on GaP NWs doping. Among them, several works consider S, Te, Sn, Se and Si as an n-type dopant while Zn seems to be the only one material considered as a p-type dopant \cite{chen2010sulfur,diak2023ultrathin,yazdi2015doping,hasenohrl2013zinc}. All the reported studies are focused on CVD-grown nanowires while to the best of our knowledge, there are no reports on MBE-grown ones. In this work, we investigate the MBE growth mechanisms of Si- and Be doped GaP NWs and quantitatively study their doping level. We optimize the growth parameters to improve the dopant incorporation and study the conductivity of individual NWs by obtaining and numerically analyzing their current-voltage characteristics (I-V curves).

\section{Results and discussion}

\subsection{Nanowire growth}
The studied GaP NWs were grown by means of molecular beam epitaxy on highly-doped (111) Si substrates. The details about the growth setup can be found in supplementary information. 

\begin{figure} [ht]
\centering
  \includegraphics[width=8cm]{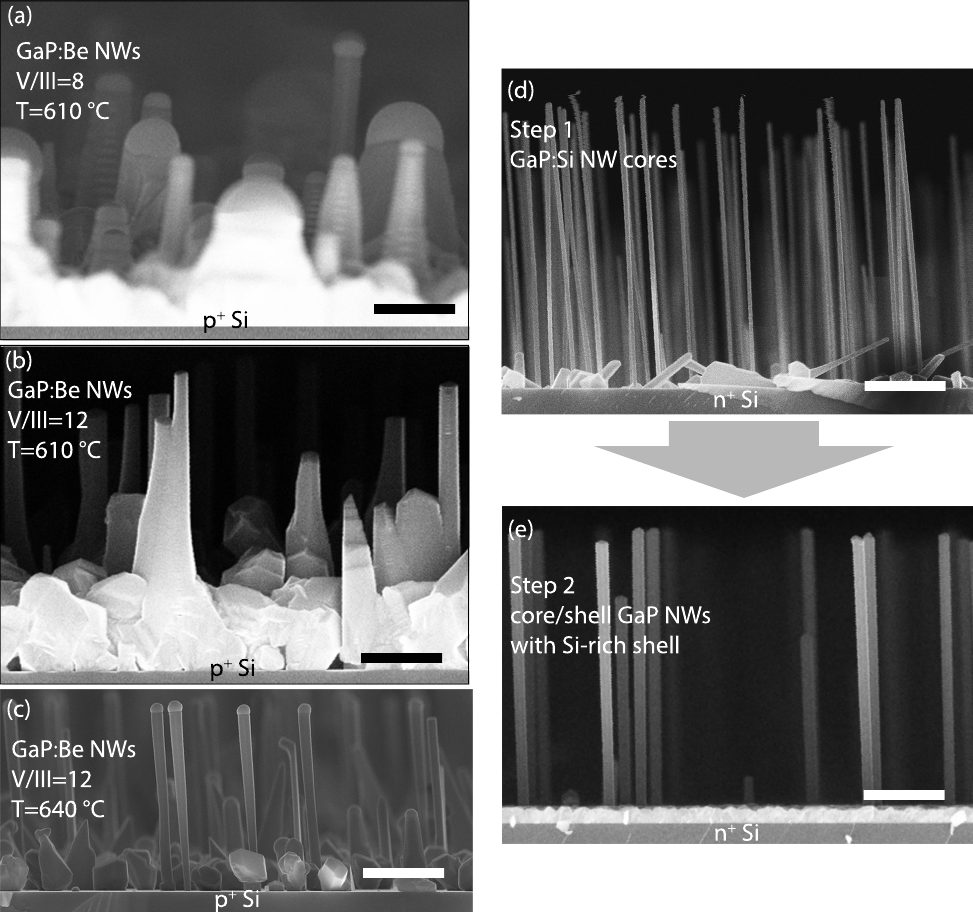}
  \caption{SEM images of as-grown GaP NW arrays: (a-c) Be-doped GaP NWs synthesized under different conditions, (c-d) two-step growth of Si-doped GaP NWs: GaP NW cores (d) and core/shell GaP NWs with Si-doped shells (e).}
  \label{SEM}
\end{figure}

According to the previous reports on the growth of Be-doped GaAs NW, the question of beryllium incorporation efficiency via the nanowire side facets vapor-solid (VS) growth \cite{Casadei_2013_GaAs_Be} or through the Ga droplet during the Ga-assisted VLS growth is debatable \cite{Zhang_2018_GaAs_Be}. However, it was shown, that regardless of the dopant incorporation path, strong lattice diffusion of Be enables homogeneous doping of NW \cite{Casadei_2013_GaAs_Be}. Our observations for the growth of GaP:Be NWs shows that at sufficiently high dopant fluxes Be accumulates in Ga droplet similarly to the case of GaAs:Be NWs reported in Ref. \cite{Zhang_2018_GaAs_Be}. In turn, Be-induced changes in the catalytic droplet morphology significantly narrow the growth window otherwise providing the uniform formation of vertical GaP NW (8 < V/III flux ratio < 30, 610 $^{\circ}$C < T\textsubscript{gr} < 630 $^{\circ}$C) \cite{Fedorov_2021}. It was found, that at high dopant fluxes Be tends to be accumulated in Ga droplet. Be-induced changes in the droplet morphology are clearly visible in Figure \ref{SEM}a,b showing SEM images of NW arrays grown under different V/III ratios. To study the catalytic droplet morphology, NW growth was interrupted by simultaneously closing P, Ga and Be fluxes. As a result, growth of GaP NW under Be flux at T\textsubscript{gr}=610 $^{\circ}$C and V/III=8 leads to the continuous Ga-droplet inflation suppressing axial NW growth and morphology changes visible on the NW side facet (see Figure \ref{SEM}a). On the other hand, as can be seen from Figure \ref{SEM}b, just 1.5-fold increase in the V/III flux ratio up to 12 leads to the absorption of the catalytic droplet interrupting VLS growth and coalescence of NWs at otherwise stable growth conditions. Thus, the range of V/III flux ratios allowing stable VLS growth turns out to be very limited, impeding the achievement of the optimal growth conditions. It was found that a stable VLS growth of GaP:Be NWs at the same Be flux can be achieved by an increase of the growth temperature T\textsubscript{gr} up to 640 $^{\circ}$C. SEM image of the Be-doped GaP NW array grown for 120 min at V/III=10 is shown in Figure \ref{SEM}c. The nanowires are 2.4 $\upmu$m long and 80-100 nm thick near the base and up to thickening to 150 $\pm$10 nm near top facet, which is covered by catalytic droplet.

In contrast to the growth of Be-doped GaP NWs, introduction of maximal achievable Si flux available in our MBE-setup (T\textsubscript{Si cell}=1200 $^{\circ}$C) does not lead to visible changes in the NW and catalytic droplet morphology. Our preliminary studies show that GaP:Si NWs demonstrate conductivity several orders of magnitude lower than expected, since the conductivity of single NW was below the sensitivity of our setup (10 pA). We associate such behaviour with amphoteric nature of silicon dopant and low probability of its incorporation on gallium sites, since self-catalytic VLS growth occurs in the excess of gallium \cite{dubrovskii2020te}.
To avoid this limitation, we introduce a two-step growth technique allowing the efficient n-type doping of GaP NWs by Si, involving preliminary formation of a thin GaP:Si NW core in the self-catalytic VLS regime followed by the formation of a highly doped GaP:Si NW shell via the VS growth. Chosen Si flux corresponded to carrier concentration in the planar GaP layer of ~7x10\textsuperscript{18} cm\textsuperscript{3}. At the first step, the NW cores are grown under Si flux at T\textsubscript{gr} of 630 $^{\circ}$C and V/III fluxes ratio of 30 for 33 min. To stop the VLS growth, a catalytic Ga droplet is consumed keeping the of P\textsubscript{2} and Si fluxes open for 20 min. As a result, a 3.8$\pm$0.8 $\upmu$m long and 60$\pm$10 nm thick NW core is formed - see Figure \ref{SEM}d. At the second step, NW shell growth is carried out at reduced growth temperature of 500 $^{\circ}$C and in excess of the group-V flux (V/III=40, which is roughly 7 times higher that stoichiometric flux ratio) for 60 min.

We suggest, that both low growth temperature and high value of group-V flux facilitate the incorporation of Si on Ga-sites contrasting to VLS growth under Ga-catalytic droplets. The obtained SEM image of the resulting GaP:Si core/shell NW array presented in Figure \ref{SEM}e confirms that radial growth prevails at the conditions chosen for the shell formation. The array possess relatively low surface density and high uniformity of NWs, which have the similar height of 4 $\upmu$m as the initial NW core and increased diameter of 160 and 130 ($\pm$ 10) nm at the top and bottom facets.

For the further investigations, the doping type of Si-doped nanowires has to be defined. Si can be considered as amphoteric dopant for GaP, thus the doping type is ambiguous. It is known that silicon can give both n- or p-type conductivity in III-V NWs \cite{hijazi2019si}. In particular, in case of GaAs NWs, it's shown that the doping type is governed by V/III flux ratio \cite{hijazi2019si}. In case of GaP NWs, Si behaviour as a dopant hasn't been studied yet. To clarify the doping type of the grown NWs, we employed photo-assisted Kelvin Probe Force Microscopy measurements under in the presence of UV illumination. The technique description and the obtained results can be found in supplementary information. The results indicate that the nanowire surface potential rises under illumination meaning that the photovoltage is positive, and therefore, Si-doped GaP NWs possess n-type conductivity.

\subsection{Conductive AFM study}
\begin{figure}[ht]
\centering
  \includegraphics[width=8cm]{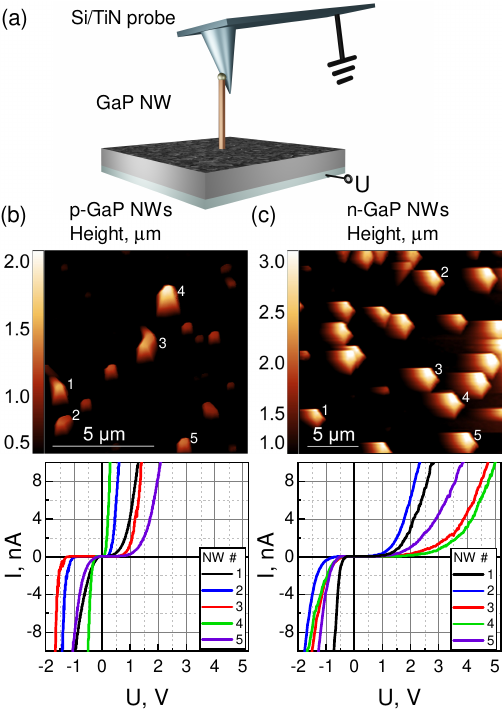}
  \caption{(a) Schematic of I-V curve measurements of an individual NW (not to scale), (b,c) surface topography and I-V curves of (b) GaP:Be and (c) GaP:Si NWs.}
  \label{experiment}
\end{figure}

The electrical conductivity of individual vertical NWs was investigated by obtaining their current-voltage characteristics (I-V curves). To do that, we used the nanopositioning technique based on conductive atomic force microscopy (C-AFM), which we have developed in several papers \cite{alekseev2015nitride,alekseev2019control,alekseev2020effect}. First, we obtained the coordinates of several NWs by scanning the sample topography. Then we switched off the microscope feed-back system and manually positioned the grounded conductive probe tip to create mechanical contact with the NW top facet. Then we applied bias voltage to the conductive substrate and obtained the current signal in the substrate/NW/probe circuit as shown in Figure \ref{experiment}a. Taking into account relatively low surface density of the grown NWs, this approach excludes shunting the electric contact with neighbouring crystals. Figure \ref{experiment}b,c shows surface topography of both studied samples as well as the obtained I-V curves for several NWs. We present the data of 5 NWs for each sample, which coordinates are indicated on topographic images with blue wheel crosses. We analyze only the currents within 10 nA in order to exclude the nonlinear effects, induced by NW heating, appearing at higher currents. 

It can be seen that the obtained curves are rather smooth indicating stable tip-NW electrical contact during the measurements. Each acquired curve is well-reproducible which was proved by conducting several dozens of consistent measurements on the same NW. The curves were obtained in dark conditions (i.e. the sample is covered with an opaque dome and the AFM laser is switched off) to avoid the photogeneration in Si substrate. All measured curves show rectifying behaviour originating mainly from NW/tip Schottky and NW/substrate contacts. The curve shape varies significantly between several NWs of each sample. We associate this with the inhomogeneity of grown arrays. Thickness and doping level naturally varies from one NW to another within the same array. Also, the curve shape depends the top electric contact area which can be slightly different in each case \cite{lord2015controlling}.

\subsection{Numerical modeling of I-V curves}
\begin{figure}[ht]
\centering
  \includegraphics[width=8cm]{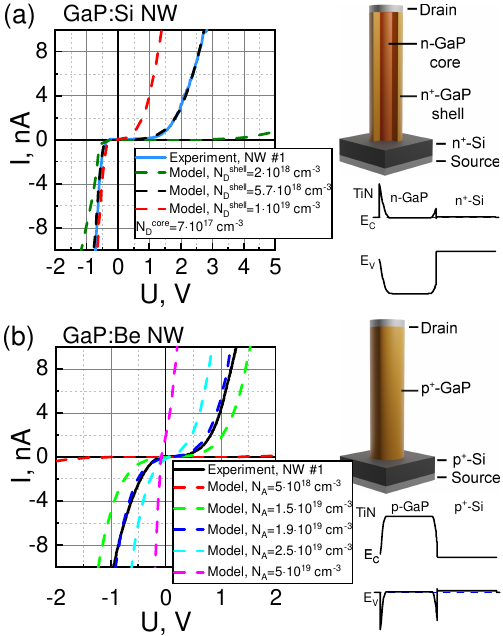}
  \caption{I-V curves modeling for (a) n-GaP NW, (b) p-GaP NW. The insets show the model schematics and energy band structure.}
  \label{AFM_VAH}
\end{figure}

To estimate the NW doping level, numerical modeling of the obtained I-V curves was performed. It's known that the curve shape is governed not only by NW intrinsic parameters such as doping concentration and carrier mobility, but also by contact and surface effects such as Schottky barrier height, contact area and surface depletion, which overall impact can change the curve shape from linear to rectifying \cite{zhang2007quantitative}. To distinguish the influence of different effects, we propose a 2D model in Silvaco Atlas software package. The NW parameters were estimated either from SEM (NW length and diameter) or from literature (GaP mobility and Schottky barrier height) and are listed in supplementary materials. We confine ourselves to modeling one curve for each sample, namely curves \#1 from the figure \ref{AFM_VAH}, and vary the doping level and work function until the best fit is achieved, considering that all the remaining curves can be modeled as well with slightly different set of parameters. The best match for the chosen experimental curves was achieved with the following parameters: N\textsubscript{D}\textsuperscript{Core}=7$\cdot$10\textsuperscript{17} cm\textsuperscript{-3}, N\textsubscript{D}\textsuperscript{Shell}=5.7$\cdot$10\textsuperscript{18} cm\textsuperscript{-3}, $\Phi_B$=1 eV. For GaP:Be NWs we have N\textsubscript{A}=1.9$\cdot$10\textsuperscript{19} cm\textsuperscript{-3} and $\Phi_B$=1.15 eV. The full set of the used values is listed in supplementary information (see table S1).

The comparison of the modeled and experimental curves is presented in Figure \ref{AFM_VAH}. The doping level appears to be the most critical parameter, which variation within one order of magnitude completely modifies the NW electrical properties. At the same time, changing the NW geometry, mobility, and Schottky barrier within the accuracy of their estimation does not lead to such dramatic changes. Impact of the doping level on the I-V curve is explained by the balance of the Schottky- and hetero-barriers connected oppositely to the NW (see band diagram in Figure \ref{AFM_VAH}). The current in the circuit substrate/NW/probe depends on the tunneling current through the barriers, because the thermionic emission is low for the high energy barriers. Since the tunneling current dominates only at high doping level, reducing the doping dramatically decreases the current (see red curve in Fig. \ref{AFM_VAH}b).

\subsection{I-V curves of horizontal GaP NW with lithographic contacts}

\begin{figure} [ht]
\centering
  \includegraphics[width=7cm]{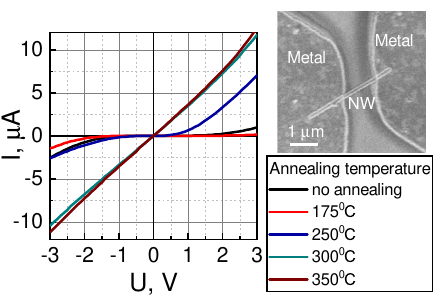}
  \caption{I-V curves of a horizontal GaP:Si NW with two lithographic metal contacts. The curves were obtained after different annealing temperatures. The inset shows SEM image of the structure.}
  \label{LITHO_SEM}
\end{figure}

C-AFM allows to study of the transport properties of numerous individual NWs which is helpful to gain statistical data and estimate their average doping level. However, the presence of Schottky barrier between the top NW facet and AFM probe complicates the analysis and can cause error. On the other hand, NW-based devices often consist of horizontally-dispersed NWs with Ohmic contacts from both sides. Thus, for validation of our numerical results and for the further development of horizontal geometry, we performed an additional independent experiment of a horizontal GaP:Si NW with two Ohmic contacts. To do that, the NW was transferred onto clean quartz substrate. After that, two metal electrodes were created using laser lithography. The creation of ohmic contacts to GaP is challenging due to the wide band gap and requires to use annealed contacts with an addition of doping metals. The main problem in case of NWs is that the annealing process can short the contacts together. To avoid this, we used the contacts based on activation process via solid phase diffusion under annealing. Based on the literature, the Pd/Si contact base is preferred to apply with NWs. The contacts were formed by depositing of Ni/Pd/Si/Al consistently with thicknesses of 3 nm, 15 nm, 30 nm, and 200 nm, respectively \cite{park1997si}. The structure was annealed after metal deposition under different temperatures from 175 to 350 $^{\circ}$C. The I-V curve obtained in such geometry is shown in Figure~\ref{LITHO_SEM} from which it follows that 350 $^{\circ}$C is sufficient annealing temperature for the contacts to become ohmic. SEM image of the NW is shown on the inset in Figure \ref{LITHO_SEM} which demonstrates no shorts between the contacts. Thus, the proposed approach is suitable for the formation of ohmic contacts to a single n-GaP NW, which is essential for NWs-based device development. To date, the formation of ohmic contacts to p-type GaP is still an issue, but we hope to implement and present it in future publications.
 
Then, specific differential resistance of the horizontal NW can be estimated as $\rho=\Delta U / \Delta I \cdot S/l$, where $l$=1800 nm and $d$=130 nm corresponds to the NW length and diameter respectively, and $S$ is a cross section area. Free charge carrier concentration is then calculated as $n=1/\rho q\mu=3.{25}\cdot10^{18}$ cm\textsuperscript{-3} which is in good agreement with the value obtained in the previous section. Therefore, we believe that the proposed numerical model predicts the doping level with good accuracy. Taking into account uncertainties in determining $l$, $d$, $\mu$, tip/NW contact area and Schottky barrier height, we can conclude that the achieved doping level lies in the range of N\textsubscript{D}=3-5$\cdot$10\textsuperscript{18} cm\textsuperscript{-3} and N\textsubscript{A}=1-2$\cdot$10\textsuperscript{19} cm\textsuperscript{-3} for GaP:Si and GaP:Be NWs respectively. Thus, the proposed growth strategies allow to synthesize samples which doping level is high enough for technological applications.

\section{Conclusion}
To conclude, in this work we synthesized bottom-up MBE-grown GaP NWs doped with Si and Be and investigated their electrical properties. We revealed a window of thermodynamic parameters facilitating a high rate of Be incorporation into GaP. We also introduced a core/shell growth algorithm allowing to use Si as an n-type dopant for GaP and successfully carried out the formation of ohmic contacts to a horizontal n-GaP NW. The I-V curves of individual NWs indicate the doping level on the order of N\textsubscript{D}=3-5$\cdot$10\textsuperscript{18} cm\textsuperscript{-3} and N\textsubscript{A}=1-2$\cdot$10\textsuperscript{19} cm\textsuperscript{-3} in case of GaP:Si and GaP:Be NWs respectively. The results open paths for the creation of axial and radial p-n junctions in GaP NWs facilitating their device implementation.

\section{Acknowledgements}

V.A.S acknowledges support of the electric measurements and numerical modeling by Russian Science foundation (Grant No. 23-72-01082). K.N.N. acknowledges support of the post-processing by Russian Science foundation (Grant No. 21-79-10346)

\section*{Conflicts of interest}
There are no conflicts to declare.

\scriptsize{
\bibliography{bibliography} 
\bibliographystyle{bibliography} }

\end{document}


\pagestyle{fancy}
\thispagestyle{plain}
\fancypagestyle{plain}{
\renewcommand{\headrulewidth}{0pt}
}

\makeFNbottom
\makeatletter
\renewcommand\LARGE{\@setfontsize\LARGE{15pt}{17}}
\renewcommand\Large{\@setfontsize\Large{12pt}{14}}
\renewcommand\large{\@setfontsize\large{10pt}{12}}
\renewcommand\footnotesize{\@setfontsize\footnotesize{7pt}{10}}
\renewcommand\scriptsize{\@setfontsize\scriptsize{7pt}{7}}
\makeatother

\renewcommand{\thefootnote}{\fnsymbol{footnote}}
\renewcommand\footnoterule{\vspace*{1pt}%
\color{cream}\hrule width 3.5in height 0.4pt \color{black} \vspace*{5pt}} 
\setcounter{secnumdepth}{5}

\makeatletter 
\renewcommand\@biblabel[1]{#1}            
\renewcommand\@makefntext[1]%
{\noindent\makebox[0pt][r]{\@thefnmark\,}#1}
\makeatother 
\renewcommand{\figurename}{\small{Fig.}~}
\sectionfont{\sffamily\Large}
\subsectionfont{\normalsize}
\subsubsectionfont{\bf}
\setstretch{1.125} 
\setlength{\skip\footins}{0.8cm}
\setlength{\footnotesep}{0.25cm}
\setlength{\jot}{10pt}
\titlespacing*{\section}{0pt}{4pt}{4pt}
\titlespacing*{\subsection}{0pt}{15pt}{1pt}

\fancyfoot{}
\fancyfoot[RO]{\footnotesize{\sffamily{1--\pageref{LastPage} ~\textbar  \hspace{2pt}\thepage}}}
\fancyfoot[LE]{\footnotesize{\sffamily{\thepage~\textbar\hspace{3.45cm} 1--\pageref{LastPage}}}}
\fancyhead{}
\renewcommand{\headrulewidth}{0pt} 
\renewcommand{\footrulewidth}{0pt}
\setlength{\arrayrulewidth}{1pt}
\setlength{\columnsep}{6.5mm}
\setlength\bibsep{1pt}

\makeatletter 
\newlength{\figrulesep} 
\setlength{\figrulesep}{0.5\textfloatsep} 

\newcommand{\topfigrule}{\vspace*{-1pt}%
\noindent{\color{cream}\rule[-\figrulesep]{\columnwidth}{1.5pt}} }

\newcommand{\botfigrule}{\vspace*{-2pt}%
\noindent{\color{cream}\rule[\figrulesep]{\columnwidth}{1.5pt}} }

\newcommand{\dblfigrule}{\vspace*{-1pt}%
\noindent{\color{cream}\rule[-\figrulesep]{\textwidth}{1.5pt}} }

\makeatother
\twocolumn[
  \begin{@twocolumnfalse}
\vspace{1em}
\sffamily

\LARGE{\textbf{Supplementary information: Electrical properties of highly-doped MBE-grown gallium phosphide nanowires}} \\

\noindent\large{Vladislav Sharov,$^{\ast}$\textit{$^{a,b}$} Kristina Novikova,\textit{$^{a,c}$} Alexey Mozharov\textit{$^{a,c}$}, Vladimir Fedorov\textit{$^{a,c}$},Prokhor Alekseev\textit{$^{b}$},Ivan Mukhin\textit{$^{a,c,d}$}} \\



 \end{@twocolumnfalse} \vspace{0.6cm}

  ]

\renewcommand*\rmdefault{bch}\normalfont\upshape
\rmfamily
\section*{}
\vspace{-1cm}


\footnotetext{\textit{$^{a}$~Alferov University, Saint-Petersburg 194021, Russia}}
\footnotetext{\textit{$^{b}$~Ioffe Institute, Saint-Petersburg 194021, Russia}}
\footnotetext{\textit{$^{c}$~Peter the Great Saint-Petersburg Polytechnic University, Saint-Petersburg 195251, Russia}}
\footnotetext{\textit{$^{d}$~Saint Petersburg State University, Saint Petersburg 199034, Russia}}
\footnotetext{\textit{$^{\ast}$~E-mail: vl\_sharov@mail.ru}}


\section{Nanowire growth}
The studied GaP NWs were grown by means of solid-source MBE using Veeco GEN III setup equipped with Si, Be and Ga effusion cells and valved phosphorous cracker cell. Beam equivalent pressures were measured using ionization pressure gauge. Ga flux was set to 8$\cdot$10\textsuperscript{-8} Torr, which corresponded to a nominal planar GaP growth rate of 180 nm/hr. Dopant fluxes were calibrated by Hall conductivity measurements of a p- and n-type doped planar GaP layers grown on i-GaP buffers on Si(001).

NW arrays were formed via self-catalyzed VLS growth mechanism assisted by Ga droplet avoiding the use of foreign metal contamination, and thus providing lower impurity level and electrically active defects \cite{Tersoff_2015}. Formation of GaP NWs occurs at the pinhole defects of the silicon surface oxide facilitating the formation of catalytic Ga droplets required for vertical NW growth. Corresponding defects in silicon oxide layer were formed \textit{in-situ} by the high-temperature annealing according to the procedure described in our previous report \cite{Fedorov_2021}. The growth of p- and n- type doped GaP NWs was carried out on highly-doped p- and n-type Shiraki-cleaned Si (111) substrates, respectively (N\textsubscript{A}\textsuperscript{Si}=1$\cdot$10\textsuperscript{20} cm\textsuperscript{-3}, N\textsubscript{D}\textsuperscript{Si}=2$\cdot$10\textsuperscript{19} cm\textsuperscript{-3}) \cite{ishizaka1986low}. NW growth was initiated by simultaneous opening of Ga, P and dopant source shutters. Axial VLS growth rate was limited by group-V flux to the catalytic droplet and varied from 20 to 120 nm/min depending on the chosen growth condition. Our observation shows that incorporation of Be and Si during NW growth is significantly different and thus two altering approaches were used to fabricate p- and n- type doped GaP NW arrays.

\section{Photo-assisted KPFM study of GaP:Si NWs}
\begin{figure}[h!]
\centering
  \includegraphics[width=6cm]{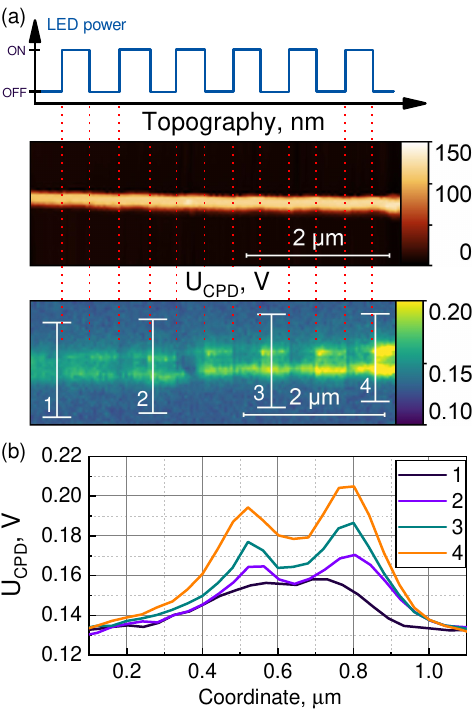}
  \caption{(a) AFM topography and KPFM surface potential map of a single Si-doped GaP NW segment under periodic UV irradiation. The upper panel shows the period of UV light switching on, (b) surface potential profiles corresponding to lines 1-4.}
  \label{figS1}
\end{figure}

Si can be considered as amphoteric dopant for GaP, thus the doping type of GaP:Si NWs is ambiguous. It is known that silicon can give both n- or p-type conductivity in III-V NWs \cite{hijazi2019si}. In particular, in case of GaAs NWs, it's shown that the doping type is governed by V/III flux ratio \cite{hijazi2019si}. In case of GaP NWs, Si behaviour as a dopant hasn't been studied yet. To clarify the doping type of the grown NWs, we employed photo-assisted Kelvin Probe Force Microscopy (KPFM) technique, which allows to measure the polarity of surface photovoltage (SPV) \cite{kronik1999surface,sharov2021work}. In our setup, contact potential is considered as $U_{CPD}=\Phi_{probe}-\Phi_{sample}$ where $\Phi_{probe}$ and $\Phi_{sample}$ is probe and sample work function respectively. We illuminate an individual horizontal NW lying on auxiliary conductive substrate with above-band-gap UV light source ($\uplambda=365$ nm). The photogenerated charge carriers change NW surface potential due to SPV ($U_{SPV}=U_{CPD}^{light}-U_{CPD}^{dark}$), which sign indicates on doping type: positive for n-type and negative for p-type \cite{kronik1999surface}. Figure \ref{figS1} shows surface topography of a 5 $\upmu$m long segment of an individual GaP:Si NW and corresponding surface potential map with 1D profiles.

UV LED was turned on periodically during the scanning in accordance with schematic plot shown in Figure \ref{figS1}a. The slow-scan direction is from left to right i.e. curve \#1 is the initial "dark" CPD profile without the influence of SPV while profiles \#2-4 are obtained under increasing UV exposure time. The map and profiles indicate that NW surface potential rises under illumination which means that $\Phi_{sample}$ decreasing due to positive photovoltage. From Figure \ref{figS1} it follows that Si-doped GaP NWs possess n-type conductivity.

\section{Numerical modeling}

The model has cylindrical symmetry and includes a GaP:Be (p-type) or GaP:Si (n-type) NW on a Si substrate. Two junctions are introduced in the model: the bottom one between the NW and substrate with ohmic behaviour and the top one between the NW and probe with Schottky barrier. GaP:Si NW is described as a core/shell structure with moderately-doped core and highly-doped shell in accordance with two-stage growth procedure. GaP:Be NW is suggested to be uniformly-doped. Thermionic emission, recombination and tunneling across the Schottky barrier are calculated using built-in universal Schottky tunneling model \cite{matsuzawa2000unified}. Thermionic emission current is calculated according to surface recombination velocity and band-to-band recombination. We use the reduced carrier mobility value $\upmu$\textsubscript{e}\textsuperscript{GaP}=$\upmu$\textsubscript{h}\textsuperscript{GaP}=10 cm\textsuperscript{2}/V$\cdot$s based on several reports \cite{hasenohrl2013zinc,montgomery1965hall,emmer2017fabrication}. The Schottky barrrier between (110) GaP and TiN is unknown so we vary the value of $\Phi_B$ around 1.23 eV which is the height of GaP/TiN barrier \cite{von1993electrical}. Silicon substrate is parameterized with the doping level, mobility and energy band parameters. The NW is parameterized with length and diameter, carrier mobility $\upmu$, GaP/TiN Schottky barrier height $\Phi_B$ and band parameters\cite{alekseev2020effect}. 
Conventionally, surface electronic states affects on the NW conductivity. However, the width of the depletion region at the NW surface non-linearly depends on NW diameter, doping level, and surface states density \cite{schmidt2007influence}. The surface states density in GaP is of 2$\cdot$10\textsuperscript{12} cm\textsuperscript{-3} and for the doping level higher the 5$\cdot$10\textsuperscript{18} the width of the depletion region is negligible \cite{lubberhuizen2000recombination}.

Below is the full set of parameters used for Silvaco modeling:

\begin{table}[h!]
\centering
\renewcommand{\arraystretch}{1.1}
\begin{tabular}{l l} 
\hline
GaP:Si NW &  \\
\hline
Length         & 4 $\upmu$m \\
Core diameter  & 30 nm      \\ 
Shell thickness & 35 nm     \\
N\textsubscript{D}\textsuperscript{Core} & 1$\cdot$10\textsuperscript{17} cm\textsuperscript{-3} \\ 
N\textsubscript{D}\textsuperscript{Shell}        & 5.7$\cdot$10\textsuperscript{18} cm\textsuperscript{-3} \\
N\textsubscript{D}\textsuperscript{Si}           & 2$\cdot$10\textsuperscript{19} cm\textsuperscript{-3} \\ 
$\Phi_B$  & 1.0 eV   \\
\\
\hline

GaP:Be NW & \\
Length         & 2 $\upmu$m \\
Diameter       & 100 nm \\
N\textsubscript{A} & 1.9$\cdot$10\textsuperscript{19} cm\textsuperscript{-3} \\
N\textsubscript{A}\textsuperscript{Si}  & 1$\cdot$10\textsuperscript{20} cm\textsuperscript{-3} \\
$\Phi_B$  & 1.15 eV  \\
\\
\hline

$\upchi$\textsuperscript{GaP} & 3.8 eV \\
$\upchi$\textsuperscript{Si} & 4.05 eV \\
E\textsubscript{g}\textsuperscript{GaP} &  2.26 eV \\
E\textsubscript{g}\textsuperscript{Si} &  1.12 eV \\
$\upmu$\textsubscript{e}\textsuperscript{GaP} & 10 cm\textsuperscript{2}/V$\cdot$s\\
$\upmu$\textsubscript{h}\textsuperscript{GaP} &  10 cm\textsuperscript{2}/V$\cdot$s\\
 \hline
\end{tabular}
\caption{Parameters in numerical model.}
\label{table}
\end{table}

\subsection{I-V curves of horizontal GaP NW with lithographic contacts}

To perform the conductive measurements in horizontal geometry, the the NW was transferred onto clean quartz substrate by ultrasonification in distilled water, followed by substrate annealing under 250 $^{\circ}$C for 2 hours. After that, two metal electrodes were created using laser lithography with bilayer polymer pattern consisted of Microchem PMGI SF9 followed by AZ MIR 701 photoresists as well as metal vacuum evaporation and lift-off process. Based on the literature, the Pd/Si contact base is preferred to apply with NWs.

\scriptsize{
\bibliography{bibliography} 
\bibliographystyle{bibliography} }